\documentstyle[amssymb,aps,prb,preprint,epsfig]{revtex}

\begin{document}
\author{\ A.E. Karakozov$^{1}$, E.G. Maksimov$^{2}$, and O.V. Dolgov$^{3}$}
\address{$^{1}$L.F. Vereshchagin Institute for High Pressure Physics,\\
Russian Academy of Sciences, 142092 Troitsk, Moscow region, Russia,\\
$^{2}$ P.N. Lebedev Physical Institute, 117924 Moscow, Russia \\
$^{3}$ Max-Planck-Institut f\"{u}r Festk\"{o}rperforschung, Heisenbergstr.\\
1, D-70506 Stuttgart, Germany}
\title{{\bf \ Electromagnetic response of superconductors and optical sum
rule.\ }}
\maketitle

\begin{abstract}
The interrelation between the condensation energy and the optical sum rules
has been investigated. It has been shown that \ the so called 'partial' sum
rule violation is related mainly to a temperature dependence of the
relaxation rate rather than to the appearance of superconductivity itself.
Moreover, we demonstrate that the experimental data on the temperature
dependence of the optical sum rule can be explained rather well by an
account of strong electron-phonon interaction.
\end{abstract}

Many recently published works are concerned with the origin of the
condensation energy of the superconducting state, a possible violation of
so-called 'optical sum rules', and the relation between these phenomena.
These papers include both theoretical investigations\cite%
{hm,nor1,scal,legg,nor2,kim} of these problems and experimental attempts\cite%
{basov,vdm,holcomb,santan} to observe a violation of the optical sum rule.
Usually, the possibility of such violation is related to the change of the
kinetic energy of metals under superconducting transition. If this statement
would be correct, than the violation of the optical sum rule should be most
clear seen in the Bardeen-Cooper-Schrieffer\cite{bcs}$(BCS)$ type
superconductors. It has been proved \ exactly by Bogolyubov\cite{bogol} that
the original model does not contain any potential energy. It is easy to see
considering the original $BCS$ Hamiltonian 
\begin{equation}
\ H_{BCS}=\sum\limits_{{\bf k},s}\varepsilon _{{\bf k}}a_{{\bf k},s}^{+}a_{%
{\bf k},s}+\sum\limits_{{\bf k},{\bf k}^{\prime }}V_{{\bf kk}^{\prime }}a_{%
{\bf k}\uparrow }^{+}a_{-{\bf k}\downarrow }^{+}a_{-{\bf k}^{\prime
}\downarrow }a_{{\bf k}^{\prime }\uparrow }\ \ .
\end{equation}%
N. Bogolyubov has proven that for the normal state 
\begin{equation}
\ \left\langle N\left| \sum\limits_{{\bf k},{\bf k}^{\prime }}V_{{\bf kk}%
^{\prime }}a_{{\bf k}\uparrow }^{+}a_{-{\bf k}\downarrow }^{+}a_{-{\bf k}%
^{\prime }\downarrow }a_{{\bf k}^{\prime }\uparrow }\right| N\right\rangle
\varpropto 1/\Omega \ \ ,
\end{equation}%
where $\Omega $ is the system volume. It means that in the thermodynamic
limit 
\begin{equation}
\ \left\langle N\left| V_{BCS}\right| N\right\rangle \equiv 0\ \ .
\end{equation}%
The $BCS$ Hamiltonian can be exactly diagonalized in the superconducting
state using the Bogolyubov-Valatin transformation 
\begin{eqnarray}
\gamma _{{\bf k}\uparrow } &=&u_{{\bf k}}a_{{\bf k}\uparrow }-v_{{\bf k}}a_{-%
{\bf k}\downarrow }^{+} \\
\gamma _{-{\bf k}\downarrow }^{+} &=&u_{{\bf k}}a_{-{\bf k}\downarrow
}^{+}+v_{{\bf k}}a_{{\bf k}\uparrow }  \nonumber
\end{eqnarray}%
It leads to 
\begin{equation}
\ H_{BCS}=\sum\limits_{{\bf k},s}E_{{\bf k}}\gamma _{{\bf k},s}^{+}\gamma _{%
{\bf k},s}\ \ 
\end{equation}%
where 
\begin{equation}
\ E_{{\bf k}}=\pm \sqrt{\varepsilon _{{\bf k}}^{2}+\Delta ^{2}},\ \ 
\end{equation}%
which is a Hamiltonian of noninteracting but superconducting quasiparticles.
The condensation energy arises from the decreasing of the ground state
eigenvalue of the expression $\left( 5\right) $ due to an appearance of a
gap $\Delta $. The same is true for any mechanisms of superconductivity. The
decrease of a properly defined one-quasiparticle energy due to the
appearance of the gap on a Fermi level is the main contribution to the
condensation energy. This phenomenon has a certain feature is common to a
metal-insulator transition, where the band structure contribution to the
total energy decreases also due to the appearance of the gap on the Fermi
level. The example of the BCS model shows that the division of the total
energy into kinetic and potential parts is not a trivial problem even for
weakly interacting quasiparticles. This division becomes even worse defined
in systems with strongly interacting electrons. The study of the optical sum
rule and its change below the superconducting transition should be based,
from our point of view, on calculations of the conductivity itself and a
detailed analysis of this function and its dependence on temperature and
frequency.

The optical sum rule can be written in general form as 
\begin{equation}
\ \int\limits_{0}^{\infty }d\omega \sigma _{1}\left( \omega \right) =\frac{%
\omega _{pl}^{2}}{8}=\frac{\pi ne^{2}}{2m}\ \   \label{1}
\end{equation}%
where $\sigma _{1}\left( \omega \right) $ is the real part of the dynamical
conductivity, $n$\ is the total electron density and $m$\ is the bare
electron mass. The function $\sigma _{1}\left( \omega \right) $ has a zero
frequency $\delta -$ function peak in the superconducting state due to the
dissipationless (rigid) response of the superconducting condensate. The
amplitude of this peak $A$ can be expressed in terms of an corresponding
penetration depth $\lambda _{L}$\ 
\begin{equation}
\ A=\frac{c^{2}}{8\lambda _{L}^{2}},  \label{2}
\end{equation}%
where $c$\ is the velocity of light . The existence of this $\delta -$
function contribution to $\sigma _{1}\left( \omega \right) $\ in the
superconducting state leads to the so-called Ferrel-Glover-Tinkham sum rule%
\cite{tinkham} 
\begin{equation}
\int\limits_{0}^{\infty }d\omega \left[ \sigma _{1}^{N}\left( \omega \right)
-\sigma _{1}^{S}\left( \omega \right) \right] =\frac{c^{2}}{8\lambda _{L}^{2}%
}\ ,  \label{3}
\end{equation}%
where $\sigma _{1}^{N,S}\left( \omega \right) $\ is the conductivity in the
normal and superconducting states, correspondingly. In such general form
this sum rule can never be violated for any superconductors possessing ideal
diamagnetic response with a finite penetration depth $\lambda _{L}$. The
real measurement of the dynamical conductivity can never be made up to
infinite frequency. They are restricted in practice to some finite value $%
\omega _{c}$. As it is well known\cite{tinkham}, the sum rule (9) is totally
satisfied in conventional superconductors when the integration is performed
up to $\omega _{c}\simeq \left( 4-6\right) \Delta $ where $\Delta $\ is the
superconducting gap. This value of $\omega _{c}$ is of the order of the
characteristic phonon energies. The statements about the sum rules
violation, which has been made in the experimental papers\cite%
{basov,vdm,holcomb,santan}, mean that the value $\omega _{c}$ in high-$T_{c}$
superconductors is much larger than in conventional ones. The maximum value $%
\omega _{c}$\ in high-$T_{c}$ systems, if they were also conventional,
should be $\simeq 0.1eV$ because they have a magnitude of the gap $\Delta
\approx 20meV.$ It has been shown in\cite{basov,basovPR} that for the
interlayer conductivity the optical sum rules are not saturated at least in
underdoped regime for $\omega _{c}\simeq 0.1eV.$ Even more intriguing
results have been obtained recently in the paper\cite{vdm,santan} where the
violation sum rules have been observed up to very high energies $\omega
_{c}\simeq 2eV.$ The main goal of the present paper is to show that the
observed violation of the optical sum rules at least for $\omega _{c}>0.1eV$
is not related explicitly to any mechanism of superconductivity. This
violation is the direct consequence of a high value of the electron
relaxation rate $\Gamma \left( \omega ,T\right) $, critical temperature $%
T_{c}$ and $\Delta $ themselves.

Let us consider the for the future discussion important the so-called
restricted or 'partial' optical sum rule. Usually it is used in the form 
\begin{equation}
\int\limits_{0}^{\omega _{c}}d\omega \sigma _{1}\left( \omega \right) =\frac{%
\pi ne^{2}}{2m_{b}}\ .\ 
\end{equation}%
Here $1/m_{b}\ $is the an effective inverse electron band mass, which is
defined as 
\begin{equation}
\frac{n}{m_{b}}=\frac{2}{\Omega }\sum\limits_{{\bf k}}\frac{\partial
^{2}\varepsilon _{{\bf k}}}{\partial k_{x}^{2}}n_{{\bf k}}\ ,
\end{equation}%
where $n_{{\bf k}}$ is an electron distribution function. For a Hamiltonian
with nearest neighbor hopping $1/m_{b}$\ can be presented in terms of a
average of the one band kinetic energy \cite{hm} 
\begin{equation}
\frac{n}{m_{b}}=\frac{a_{x}^{2}}{\Omega }\left\langle -T_{kin}\right\rangle
\ ,\ 
\end{equation}%
\begin{equation}
\ T_{kin}=-\sum\limits_{i}t_{i,i+a_{x}}a_{i}^{+}a_{i+a_{x}}\ .\ 
\end{equation}%
Here $a_{x}$\ is the lattice spacing in $x$\ - direction and $t_{i,i+a_{x}}$
is the nearest neighbor hopping integral. Usually it is believed that the
high energy cut off \ frequency $\omega _{c}$\ should be chosen of the order
of the corresponding band plasma frequency $\tilde{\omega}_{pl}=\sqrt{4\pi
e^{2}n/m_{b}}$ and is much smaller than the energies of interband
transitions. This partial sum rule can be easily proved for noninteracting
band electrons including the {\it interband} transitions in the expression
for the conductivity 
\begin{equation}
\sigma _{1}^{inter}\left( \omega \right) =\frac{2\pi e^{2}}{\Omega m^{2}}%
\sum\limits_{{\bf k},j}n_{{\bf k}}\frac{\left| \left\langle {\bf k}j\left|
\nabla _{x}\right| {\bf k}\right\rangle \right| ^{2}}{\varepsilon _{{\bf k}%
j}-\varepsilon _{{\bf k}}}\delta \left( \varepsilon _{{\bf k}j}-\varepsilon
_{{\bf k}}-\omega \right) \ \ .
\end{equation}%
Here the summation is over all empty high energy bands with an electron
dispersion $\varepsilon _{{\bf k}j}.$ The minimum of the value $\varepsilon
_{{\bf k}j}-\varepsilon _{{\bf k}}=E_{g}$\ is the energy of the interband
transitions. Now, using $\omega _{c}<E_{g}$\ \ we can easily prove the sum
rules $\left( 10\right) $ and $\left( 12\right) $\ which give the well known
identity for the electron inverse effective mass\cite{AM} 
\begin{equation}
\frac{n}{m_{b}}=\frac{n}{m}-\frac{2}{\Omega m^{2}}\sum\limits_{{\bf k},j}n_{%
{\bf k}}\frac{\left| \left\langle {\bf k}j\left| \nabla _{x}\right| {\bf k}%
\right\rangle \right| ^{2}}{\varepsilon _{{\bf k}j}-\varepsilon _{{\bf k}}}\
\ .
\end{equation}%
However is it not the case for interacting electrons. This fact can be
easily understood using the model of electrons interacting with impurities.
The {\it intraband} contribution to the optical conductivity can be written
in this case in form of the usual Drude expression for $\sigma _{1}\left(
\omega \right) $

\begin{equation}
\sigma _{1}\left( \omega \right) =\frac{\tilde{\omega}_{pl}^{2}}{4\pi }\frac{%
\Gamma }{\omega ^{2}+\Gamma ^{2}},
\end{equation}%
where $\Gamma /2$\ is the relaxation rate due to impurity scattering. One
can derive the well known result for the partial optical sum rule

\begin{equation}
\int\limits_{0}^{\omega _{c}}d\omega \sigma _{1}\left( \omega \right) =\frac{%
\tilde{\omega}_{pl}^{2}}{8}\left( 1-\frac{2\Gamma }{\pi \omega _{c}}\right) .
\end{equation}%
This example shows that the {\it intraband} sum rules $\left( 10\right) $\
and $\left( 12\right) $ can be satisfied in the presence of the interaction
only in the limit $\omega _{c}\rightarrow \infty $. It is also true for any
interactions other than impurity scattering including, for example, the
electron-phonon interaction. Moreover, this violation of the optical sum
rules can not be obtained from the calculation of the kinetic energy change
from $Eq.\left( 12\right) $ as it was made in the Refs.\cite{hm,nor2}. As it
follows from $Eq.\left( 17\right) $\ the optical sum rules violation depends
on the high energy cutoff $\omega _{c}$\ but this parameter is certainly
absent in the expression for the kinetic energy.

The {\it interband} transitions will also be changed due to electron
interactions. We can rewrite $Eq.\left( 14\right) $ for the impurity
scattering model in the simplest approximation as

\begin{equation}
\qquad \sigma _{1}^{inter}\left( \omega \right) =\frac{2\pi e^{2}}{\Omega
m^{2}}\sum\limits_{{\bf k},j}n_{{\bf k}}\frac{\left| \left\langle {\bf k}%
j\left| \nabla _{x}\right| {\bf k}\right\rangle \right| ^{2}\Gamma }{\left(
\varepsilon _{{\bf k}j}-\varepsilon _{{\bf k}}-\omega \right) ^{2}+\Gamma
^{2}}.\qquad \ 
\end{equation}%
The general sum rule $\left( 7\right) $ is certainly satisfied in this model
for any value of the relaxation rate $\Gamma $ but it is not true for the
partial sum rule as we have discussed above. Further, as it follows from $%
Eq.\left( 18\right) ,$ the {\it interband \ }contribution to the
conductivity becomes now spread out over all intervals of energies including
very low $\omega .$ It means that we can not even divide experimental data
into {\it intraband} contributions and {\it interband} ones. All these
effects are small as $\Gamma /\omega _{c}$ and $\Gamma /E_{g}.$ However, if
we shall take into account that the relaxation rate in high-$T_{c}$
compounds can reach values $\Gamma \approx 100meV$ we see immediately that
these effects can be very important even in the normal state.

As the discussion given above confirms, there is no other way to understand
the behavior of the partial sum rules than to calculate the conductivity
itself as a function of frequency, temperature, doping etc. It can be
obtained by a calculation of a current-current correlation function. We
would like to emphasize here that expression for the current-current
correlation function does not contain, at least in the absence of the vertex
corrections, any explicit information about the mechanism of
superconductivity. All implicit information about the mechanism is contained
in the one-particle Green's function which can be written as\qquad 
\begin{equation}
\hat{G}^{-1}\left( {\bf k},\omega \right) =\omega Z\left( {\bf k},\omega
\right) \hat{1}-\varepsilon _{{\bf k}}\hat{\tau}_{3}-Z\left( {\bf k},\omega
\right) \Delta \left( {\bf k},\omega \right) \hat{\tau}_{1}.  \label{8}
\end{equation}%
Here $Z\left( {\bf k},\omega \right) $ is a renormalization function, $%
\Delta \left( {\bf k},\omega \right) $ is the superconducting order
parameter, and $\hat{\tau}_{i}$ are Pauli matrices. These functions should
be calculated in turn from the general equations for the Green's function of
electrons. Such equations have been derived by Eliashberg\cite{eliash} for
conventional superconductors with the electron-phonon pairing mechanism and,
for example, in Refs.\cite{mbp} for $d-$wave superconductivity. The
expression for $\sigma ^{S}\left( \omega \right) $ can be written for $%
\omega \gg 2\Delta $ in the form 
\begin{eqnarray}
\frac{\sigma _{1}^{S}\left( \omega \right) }{\sigma _{1}^{N}\left( \omega
\right) } &=&\frac{2}{\omega }\int\limits_{\Delta }^{\omega /2}d\omega
^{\prime }\{%
\mathop{\rm Re}%
\frac{\omega -\omega ^{\prime }}{\sqrt{\left( \omega -\omega ^{\prime
}\right) ^{2}-\Delta ^{2}\left( \omega -\omega ^{\prime }\right) }}%
\mathop{\rm Re}%
\frac{\omega ^{\prime }}{\sqrt{\omega ^{\prime 2}-\Delta ^{2}\left( \omega
^{\prime }\right) }}-  \label{9} \\
&&\ \ \ 
\mathop{\rm Re}%
\frac{\Delta \left( \omega -\omega ^{\prime }\right) }{\sqrt{\left( \omega
-\omega ^{\prime }\right) ^{2}-\Delta ^{2}\left( \omega -\omega ^{\prime
}\right) }}%
\mathop{\rm Re}%
\frac{\Delta \left( \omega ^{\prime }\right) }{\sqrt{\omega ^{\prime
2}-\Delta ^{2}\left( \omega ^{\prime }\right) }}\}.\qquad  \nonumber
\end{eqnarray}%
The expression on the right hand side of the $Eq.\left( 20\right) $ is
nothing else than the BCS type coherency factors. The $Eq.\left( 20\right) $
have been derived\cite{nam,ss} for the conventional superconductors and was
used recently\cite{holcomb} to discuss the problem of the influence of the
superconducting gap on the optical spectra at $\omega _{c}\simeq 1.2eV$. In
spite of that this expression has been derived in the framework of the usual 
$s-$wave superconductivity, it has with a slight modification much wider
areas of applications. One can, for example, include the angular dependence
of the gap for anisotropic superconductors and perform the integration over
the angle. It is easy to show using $Eq.\left( 20\right) ,$ that for $\omega
\gg 2\Delta $ we have

\begin{equation}
\sigma _{1}^{S}\left( \omega \right) =\sigma _{1}^{N}\left( \omega \right)
\left( 1-\alpha \frac{\Delta ^{2}}{\omega ^{2}}\right) .  \label{10}
\end{equation}%
Here the numerical coefficient $\alpha $ is of the order of unity and it is
included to take into account the possible averaging of the angular
dependence of the gap function. The same estimation for the dynamical
conductivity of a superconducting state at $\omega \gg \Delta $ can be
obtained from the equations derived in Ref.\cite{mbp}. $Eq.\left( 21\right) $
shows that the direct contribution of the superconducting gap to the
dynamical conductivity and, therefore, to the optical sum rules has the same
smallness for any mechanism of superconductivity, that is $\left( \Delta
/\omega \right) ^{2}$. This smallness is, certainly, different for
conventional superconductors and high-$T_{c}$ ones because the gap in the
later is one order larger.

$Eqs.\left( 14\right) ,\left( 18\right) $ for the {\it interband}
contribution can also be generalized for the superconducting case and it can
be shown that their difference from the normal state is of the order of $%
\left( \Delta /\omega \right) ^{2}.$ We shall not consider the behavior of
the optical sum rules in the superconducting state further in this paper
because the real mechanism of superconductivity in high-$T_{c}$ systems is
unknown. The investigation of the optical sum rules for the normal state
will be of our main interest in the rest part of this paper. The detailed
experimental study of this problem has been done by D. van der Marel and
coworkers\cite{vdm}. They have measured the conductivity in a wide frequency
range and temperature intervals and than two optical sum rules have been
calculated. One of them was the low energy sum rule $A_{L}$

\begin{equation}
A_{L}=8\int\limits_{0^{+}}^{1.25eV}d\omega \sigma _{1}\left( \omega \right) +%
\frac{c^{2}}{\lambda _{ab}^{2}\left( T\right) }\ 
\end{equation}%
where $\lambda _{ab}$ is the penetration depth in the $CuO$ plane and the
other one is the high energy sum rule $A_{h}$

\begin{equation}
A_{h}=8\int\limits_{1.25eV}^{2.5eV}d\omega \sigma _{1}\left( \omega \right) .
\label{11}
\end{equation}%
A few very prominent features in the behavior of both $A_{L}$ and $A_{h}$%
have been found in this work. First, $A_{L}$ and $A_{h}$ are temperature
dependent in the superconducting state as well as in the normal one. Second,
this dependence, at least, in the normal state is well described by a
quadratic function of $T.$ In addition, the low energy part depends on the
high-energy cutoff frequency $\omega _{c},$ if we consider the cases $\omega
_{c}=1.25eV$ and $\omega _{c}=2.5eV.$

We are coming now to the consideration of details in the high energy part of
the optical sum rule $A_{h}.$ Preliminary we shall neglect the direct
contributions of\ the superconducting gap to the value of $A_{h}$ because
the ratio $\Delta ^{2}/\omega ^{2}$ for the considered values of frequencies
is very small $\approx 2\cdot 10^{-4}$. The main problem in the calculation
of the normal state conductivity is to establish the origin of the electron
relaxation in high-$T_{c}$ systems. This problem along with the problem of
the origin of superconductivity itself has been disputed during the last 15
years. It was shown (see for details\cite{maksUFN}) that the main source of
the relaxation processes in the normal state of high-$T_{c}$ superconductors
is the strong electron-phonon interaction. Recently, it has been
additionally demonstrated through examination of the frequency and
temperature dependence of the optical reflectivity in the $YBCO$ system\cite%
{maksSSC} that this interaction leads to a strong temperature dependence of
the conductivity up to very high frequencies. The experimental verification
of the existence of strong electron-phonon interaction in high-$T_{c}$
superconductors has been also obtained in ARPES measurements\cite{lanzara}
as an effect of an electron mass renormalization. There is some discussion%
\cite{kee,abanov} about the possibility, that the electron mass
renormalization observed in Ref.\cite{lanzara}and the corresponding change
of the relaxation rate can been explained by the interaction with the
so-called 'magnetic resonance peak'. This possibility, however, is unlikely%
\cite{kee,shen}, at least, for the normal state. As is well known, in the
normal state the conductivity $\sigma ^{N}\left( \omega ,T\right) $ in a
presence of the strong electron-phonon interaction can be written in the form%
\cite{sdm,maksUFN}

\begin{equation}
\sigma ^{N}\left( \omega ,T\right) =\frac{\omega _{pl}^{2}}{4\pi }\frac{1}{%
-i\omega \frac{m^{\ast }\left( \omega ,T\right) }{m}+\Gamma \left( \omega
,T\right) },  \label{12}
\end{equation}%
where $m^{\ast }\left( \omega ,T\right) $ is the frequency dependent optical
mass and $\Gamma \left( \omega ,T\right) $ is the optical relaxation rate.
The readers can find the precise expressions for both these functions in
terms of the Eliashberg function in Refs.\cite{sdm,maksUFN} and we shall not
reproduce them here. $Eq.\left( 24\right) $ for high values of frequencies $%
\left( \omega \gg \{\omega _{ph},\Gamma \}\right) $ can be rewritten for the
real part of the conductivity in the form

\begin{equation}
\sigma _{1}^{N}\left( \omega ,T\right) \approx \frac{\omega _{pl}^{2}}{4\pi }%
\frac{\Gamma \left( T\right) }{\omega ^{2}},  \label{13}
\end{equation}%
where $\Gamma \left( T\right) $ is independent on frequency\cite{sdm,maksUFN}

\begin{equation}
\Gamma \left( T\right) =2\pi \int\limits_{0}^{\infty }d\Omega \alpha
_{tr}^{2}\left( \Omega \right) F\left( \Omega \right) \coth \frac{\Omega }{2T%
}.  \label{33}
\end{equation}%
Here $\alpha _{tr}^{2}\left( \Omega \right) F\left( \Omega \right) $ is the
transport Eliashberg function. It is easy to show by using $Eq.\left(
26\right) $, that 
\begin{equation}
\Gamma \left( T=0\right) =\lambda _{tr}\pi \left\langle \omega \right\rangle
,  \label{34}
\end{equation}%
where

\begin{equation}
\lambda _{tr}=2\int\limits_{0}^{\infty }d\Omega \frac{\alpha _{tr}^{2}\left(
\Omega \right) F\left( \Omega \right) }{\Omega }  \label{28}
\end{equation}
is the {\it transport } constant of EPI, and $\left\langle \omega
\right\rangle $ is the average phonon frequency. At considerably high
temperatures, on the other hand, $\Gamma \left( T\right) $ can be written as

\begin{equation}
\Gamma \left( T\right) \approx 2\lambda _{tr}\pi T.  \label{16}
\end{equation}%
$Eqs.\left( 27\right) $ and $\left( 29\right) $ show that the relaxation
rate can increase considerably with increasing of temperature. It will lead,
to some increase of the high frequency part of the optical sum rule, which
can be written in the form

\begin{equation}
A_{h}=8\int\limits_{\omega _{1}}^{2\omega _{1}}d\omega \sigma _{1}(\omega )=%
\frac{\omega _{pl}^{2}}{\pi }\frac{\Gamma (T)}{\omega _{1}},  \label{17}
\end{equation}%
where $\omega _{1}=1.25eV$. Using $Eqs.(25)$ and $(26)$ we can easily
calculate this value. There are two independent fitting parameters in this
procedure: the plasma frequency $\omega _{pl}$ and the coupling constant $%
\lambda _{tr}$. We have chosen the value $\lambda _{tr}\lesssim \lambda
\approx 1.5$ in accordance with APRES data \cite{lanzara}. The value of the 
{\it intraband} plasma frequency is also unknown, but it is bounded from
above by the value of the low part of the sum rule obtained in Ref.\cite{vdm}%
, that is

\begin{equation}
\omega _{pl}\lesssim 2eV.  \label{18}
\end{equation}

For the numerical calculations we have employed the Eliashberg function from
our preceding papers\cite{sdm,maksUFN} and use the general expression for
the conductivity $\left( 24\right) ,$ rather than approximate $Eq.\left(
25\right) $. We carried out our calculations for two slightly different
Eliashberg functions shown in the inset in Fig.1, having the same value of $%
\lambda _{tr}.$ The difference between these spectra is related to the
different coupling of electrons with a soft phonon $\omega _{ph}\thickapprox
20meV)$ and harder ones. The temperature dependence of $A_{h}$ is shown in
Fig.1 at $0\lesssim T\lesssim 200K.$ We have used as in Ref.\cite{vdm} the $%
T^{2}$ scale for the temperature to demonstrate the near perfect quadratic
dependence $A_{h}$ on $T.$ The overall agreement of our results with the
experimental data is reasonably well. The same is true concerning the
experimentally observed difference 
\begin{equation}
A_{h}\left( T=200K\right) {\bf -}A_{h}\left( T=0K\right) {\bf \thickapprox }%
0.08\left( eV\right) ^{2}.
\end{equation}%
It can also be seen from Fig.1, that $A_{h}$ has a temperature dependence
also at $T<T_{c},$ where $T_{c}$ is the critical temperature of the
superconducting transition ( $T_{c}=88K$ for the considered case).{\bf \ }We
would also like to emphasize that there is a little different behavior of
these curves at low temperatures. The value $A_{h}$ decreases with
decreasing of temperature for a softer spectrum even faster for low
temperature than $T^{2}$ as it has for high $T$. In contrast, $A_{h}$ has a
more weaker temperature dependence at low $T$ for more harder spectrum. The
difference $A_{h}\left( T=200K\right) {\bf -}A_{h}\left( T=0K\right) $ for
the spectrum with soft low frequency phonons is larger then for the harder
one.{\bf \ }We did not take into account in our calculations the influence
of the superconductivity on $A_{h}.$ It is small from our point of view but
it can exist. It is clear from the above consideration that it is very
difficult to separate using experimental data this specific superconducting
contribution from the total change of $A_{h}$ connected with the change of
the relaxation rate.{\bf \ }

The general behavior of the low energy sum rule $A_{L}\left( T\right) $ (not
shown in Fig.1) is also reproduced rather well in our approach, at least for
the normal state. There is only one contradiction related to the total
amplitude of the change of the value $A_{L}\left( T\right) .$ It is clear
from the above consideration that the following equality should be satisfied
in the normal state

\begin{eqnarray}
A_{L}\left( T=200\right) -A_{L}\left( T=100\right) &=&A_{h}\left(
T=200\right) -A_{h}\left( T=100\right)  \nonumber \\
&&+A_{h}^{\prime }\left( T=200\right) -A_{h}^{\prime }\left( T=100\right) ,
\end{eqnarray}%
where

\begin{equation}
A_{h}^{\prime }=8\int\limits_{2.5eV}^{\infty }d\omega \sigma _{1}\left(
\omega \right) .\ 
\end{equation}%
It is easy to see that

\begin{equation}
A_{h}^{\prime }=A_{h}.\ 
\end{equation}%
It means that the total change of the low energy sum rule $A_{L}\left(
T\right) $ should be twice larger than the change of $A_{h}\left( T\right) $%
. Measurements\cite{vdm} give rather the value $1.5$ instead of two. We do
not know the exact origin of this contradiction. It is possible that it is
related to the temperature dependence of the {\it interband} transitions
which have not been included into our calculations. Indeed, we have obtained
as the value of the intraband contribution to $A_{h}{\bf \approx }0.21\left(
eV^{2}\right) $ at $T=200K$. It is much smaller than the experimentally
measured $A_{h}$\ $\thickapprox 1.8\left( eV\right) ^{2}.$ The difference
between these two values comes from the {\it interband} transitions. It is
difficult to say anything definite at this time about the temperature
dependence of the {\it interband} transitions and we will continue our
activity in this direction. Now we would like to emphasize that it not easy
to find any other mechanism of the relaxation besides the electron-phonon
one which can lead to a temperature dependence of the relaxation rate at so
high frequencies. Many of them, including, for example, the marginal Fermi
liquid\cite{varma}, do not give any temperature dependence at $T\ll \omega .$

We can not calculate and compare with the experiment the sum rules in the
underdoped regime due to the existence of a {\it pseudogap} phenomenon
because its origin is also unknown. We can, however, claim that it is very
likely that the discussed effect will be also exist in the underdoped case.
This statement is based on observations obtained both by optical
measurements \cite{puchkov} and as well as ARPES\cite{lanzara}, that the
relaxation rate increases with decreasing the doping level. This also is
confirmed by result obtained in Ref.\cite{santan} on the sum rules violation.

In summary, we have shown that there are no new energy scales defining the
influence of the superconductivity on the intraband contribution to the
optical sum rules, besides the superconducting energy gap itself. This is
true for any mechanism of the superconductivity because the expression for $%
\sigma _{1}\left( \omega ,T\right) $ at high frequencies does not include
any explicit information about such mechanisms.\ The experimentally observed
violation of the sum rule is mainly related to the properties of the normal
state of high-$T_{c}$ superconductors and it is ruled mainly by the
frequency and temperature dependence of the relaxation rate. We also have
shown that the experimental data obtained in the Ref.\cite{vdm} can be
explained very reasonably in the framework of the usual model with strong
electron-phonon interaction. The consideration of the sum rules for the
interplanar conductivity where the coherent transport is absent in the
normal state requires a more serious approach and the knowledge of the
mechanism blocking this transport. We should also know more details about
the pseudogap phenomenon and its interplay with superconductivity in order
to make more conclusive statements about the sum rule behavior in the
underdoped regime.

\bigskip

Acknowledgments: We are very grateful to many people for the fruitful
discussion and especially to N. Bontemps, A. Boris, C. Bernhard and D. van
der Marel. We would like to thank J. Kortus for his help in the preparation
of the paper. This work was supported partially by RFBI under grants
No.02-02-16658 and No.01-02-16719, the Russian program for the investigation
of the superconductivity and the program of the Presidium RAS. One of the
authors (E.G.M.) is grateful to O.K. Anderson for the kind hospitality
during the visit in the Max-Planck-Institute FKF (Stuttgart) where the part
of this work has been done.

\bigskip \newpage 

\section{Figure caption.}

\bigskip 

Fig.1 $A_{h}$ (see text) as a function of $T^{2}$ for two different spectral
functions as  shown in the inset.

\qquad \qquad \qquad \qquad \qquad

\bigskip

\qquad

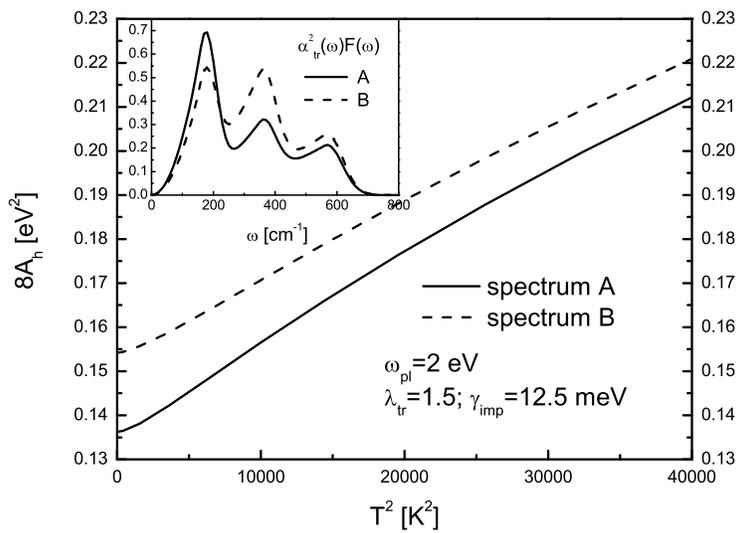
\end{document}